\documentclass[aps,preprint,showpacs,amsmath,amssymb,superscriptaddress]{revtex4-1}

\usepackage{graphicx}
\usepackage{color}

\begin{document}
\title{Gravitational waves in Brans-Dicke Theory with a cosmological constant}

\begin{abstract}
Weak field gravitational wave solutions are investigated in Brans--Dicke (BD) theory in the presence of a cosmological constant. In this setting  the background geometry is   not   flat but asymptotically de-Sitter. We  investigate the linearised field equations, and their gravitational wave solutions in a certain gauge choice. We will show that this theory leads to  massless scalar waves as in original BD theory and in contrast to massive BD theory.  The effects of these waves on free particles and their polarization properties are studied extensively and effects of the cosmological constant is analyzed in these phenomena in detail. The energy  flux of these waves are also discussed in this background. By analyzing this flux,  we obtain a critical distance where the waves cannot propagate further, which extends Cosmic no Hair Conjecture (CNC) to BD theory with a cosmological constant. 

\end{abstract}


\author{Hatice \"Ozer}
\email{hatice.ozer@istanbul.edu.tr}
\affiliation{Department of Physics, Faculty of  Sciences,  Istanbul University, 34134  Istanbul, Turkey}
\author{\"Ozg\"ur Delice}
\email{ozgur.delice@marmara.edu.tr}
\affiliation{Department of Physics, Faculty of  Arts and Sciences, Marmara University, 34722  Istanbul, Turkey}
\date{\today}

\maketitle

\section{Introduction}

 Einstein's theory of General Relativity (GR)  is a  very successful theory of gravitation which perfectly explains all related phenomena and passes all of the tests in the weak gravity regime \cite{MTW}.  Its predictions on strong field phenomena, such as on cosmology and black holes,  opened new windows on understanding the structure of the universe. Despite these achievements,  the research on its alternative theories does not seem to come to an end and they were getting  a lot of attention in last years \cite{Sotiriou,Nojiri,Capozziello,Clifton}. The motivations of these alternative theories have   several different reasons. First of all, in order to understand the mathematical structure and physical predictions of general relativity, its alternative theories should be studied. In this context, we can make modifications in GR and can study mathematical and physical consequences of these modifications. Then, we can compare the predictions of GR and   its alternative theories and determine the conditions whether these theories could be compatible with the available observational data. Another motivation comes from  the attempts to quantize gravity, which requires higher order modifications on the Einstein--Hilbert (EH) action and  indicate deviations from GR \cite{Utiyama,Stelle}. Some motivation comes from dark components, namely the dark matter and the dark energy, of the matter-energy composition of the Universe. These components were included  to make GR compatible with observations on intergalactic and cosmological scales. Although these dark components might possibly be effects of yet to be discovered particles which might be  observed in the scheme of standard model or beyond, they certainly imply that it may be worthwhile to investigate the possibility that these dark components are just the effects caused by the modifications on large scales of GR.  One last motivation we can list is the unification of gravity with other forces, which requires the modifications on EH action \cite{Lidsey}, as in the Kaluza--Klein  theory \cite{Kaluza,Klein}.

 After the discovery of GR by Einstein, one interesting prediction was made by Einstein himself in 1916 \cite{Einstein}  that the fabric of space-time could ripple. Namely, he predicted the existence of gravitational waves, tiny propagating ripples in the curvature of  space-time. This topic becomes  one of the most important topics of GR, together with black holes and cosmology. Although an indirect evidence is observed    decades ago \cite{indirect },  their interaction with a matter distribution when passing through it is extremely tiny that their direct observation is required a century to be passed after their prediction by Einstein.   In order to detect gravitational waves directly,  very sensitive devices were built such as  laser interferometric gravitational wave antennas LIGO and Virgo. The first direct observation of gravitational waves \cite{detection} was made on 14 September 2015 by LIGO antennas. The source of these waves is the catalystic event  that the coalescence and merger of two black holes of 65 $M\odot $ and 22 $M\odot$ \cite{properties}. After this observation, many more gravitational waves were observed by LIGO and VIRGO collaborations corresponding to  coalescence of black hole binaries, neutron star binaries \cite{Neutron}  in the first, second \cite{GWTC-1} and third \cite{GWTC-2} observing runs. All these observations are a living proof that the gravitational wave observations are opened a new window to the universe.  The GW observations are compatible with GR predictions \cite{test,test1,test2} which  may also help to rule out or limit corresponding predictions of alternative gravity theories within the limits of the detectors. Hence, it is important to understand the predictions of the alternative theories \cite{Sotiriou,Nojiri,Capozziello, Clifton} about properties of gravitational waves \cite{Will} 
 in their framework to estimate their viability as alternatives to GR.  

 One of the most simple and most studied alternative theory of gravity is Brans-Dicke (BD) scalar tensor theory \cite{Brans,BransDicke}. In this theory, the gravitational interaction is mediated by both the curvature of the spacetime represented by a non flat metric tensor and a also scalar field, which takes the role of Newton's gravitational constant.  In the BD theory in its original form, the extra scalar field is a long range one. This property, together with observational results that sets the free parameter of this theory, the so called BD parameter, $\omega $, to very high values, $\omega>40000 $ \cite{Bertotti}, which makes BD theory indistinguishable from GR.  In order to overcome this, a potential term, which brings an effective mass to the scalar and makes the scalar field a short range one, is added in some works. This modification of the theory is called as BD theory with a potential ( let say BDV theory), or massive BD theory. This arbitrary potential effectively plays the role of a cosmological constant in BD theory. However, adding an arbitrary potential to the action is not the only way of bringing an effective cosmological "constant" to this theory. It is possible to extend BD theory by adding a cosmological constant term whose coupling with the scalar field is exactly the same as its coupling to the Ricci curvature scalar in original BD  action. This theory is known as BD theory with a cosmological constant (BD$\Lambda$ theory). The main difference between these extensions of BD theory is that, in the latter case, the scalar field is still long range one. Hence, BD$\Lambda$ theory keeps the spirit of original BD theory by having a long range scalar field even in the presence of cosmological constant.
 
 BD$\Lambda$ theory has important consequences in different areas which makes it worthwhile to study. In cosmology, the presence of cosmological constant gives much larger solution space then GR and BD theories, allowing some solutions such as inflationary  perfect fluid \cite{Pimentel,Kolithic} or vacuum \cite{Romero,Romero1} solutions not possible in the original BD theory. Another important class of such solutions are ``the bounce solutions" \cite{Kolithic,Tretyakova:2011ch} where the time reversal of cosmological expansion does not lead to a big bang singularity.  In this case, the scale factor becomes finite and never vanishes during its evolution backwards in time, leading to a singularity free cosmology.  The extra degrees of freedom due to presence of cosmological constant term also gives other interesting  solutions and  results not present in BD theory, such as,  interior static cosmic string solutions \cite{delice1} satisfying cosmic string equation of state where original BD theory is not compatible with, existence of  Gödel type solutions in BD$\Lambda$ theory \cite{Aguledo} which are not present in BD theory,  a possible extension of Horava-Lifshitz gravity to BD theory with the help of a negative cosmological constant \cite{Oh}  in BD$\Lambda$ theory.  Some other applications of BD$\Lambda $ theory can be found  in the works \cite{Lorenz,Lorenz1,Uehara,Barrow,Pandey,Novikov,delice2,baykaldeliceciftci}.   Recently, local weak field effects of BD$\Lambda$  theory were  discussed comparatively with BDV theory in a recent paper \cite{Ozer} to see their implications and differences of these theories in that regime.

  In this paper we want to explore the effects of the presence of a cosmological constant on the properties of gravitational waves in BD$\Lambda$ theory such as their interaction with matter fields during their propagation in the spacetime, their polarization states  and their energy flux. In General Relativitiy, the gravitational waves propagate at the speed of light and they posses two independent polarization states: the plus and cross modes, which were compatible with LIGO results. Note that gravitational waves in massive BD theory is a well known topic \cite{Alsing,Maggiore,Will} hence we refer these works on gravitational waves in massive BD theory. 
 However in those works the effects of the minimum of the potential is ignored to have an asymptotically flat spacetime.

  We will first  discuss the known solution of  weak field equations of Brans-Dicke  scalar tensor theory corresponding to a massive point particle in different coordinates, which can be obtained by appropriate coordinate transformations. Then we will determine  asymptotically de-Sitter background  geometry by solving the linearised vacuum solutions of these specific theories. Finally, we will obtain the linearised gravitational wave solutions of these equations in the relevant order of the background parameters. With the help of these new solutions, we will analyse the propagation of gravitational waves in an asymptotically de-Sitter space for these scalar tensor theories and then we will find the polarization states of gravitational waves and compare the results with the results found in GR theory. Finally, we will calculate the energy flux of these waves using the short wave approximation method. 
  
  The paper is organized as follows. In Section (II) we will discuss 
  the linearised field equations in the Lorentz gauge for BD$\Lambda$  theory.
   Then we will briefly consider a static point mass solution in static isotropic coordinates to show that scalar field has long range.
    In section (II)  the linearised gravitational wave solutions will be obtained for this theory. The detection of gravitational waves by LIGO  showed that the properties of observed waves  do not have any conflict with these waves in  general relativity. However  these detectors cannot detect by design  the scalar breathing modes and the BD theory is not ruled out.   In order to construct detectors capable of detecting gravitational waves of these theories, their physical properties such as their polarizations and speed of propagation must be determined.  Hence in section (III A and B), we obtain background and wave solutions. In section (III.C) we  study their effects on free particles, study their polarizations and  propagation speeds  of the polarization modes in the BD$\Lambda $ theory. Comparison of these properties with some other theories is also made in Section (III.D).  In the Section (IV)  the energy-momentum tensor of gravitational waves for the BD$\Lambda $ theory are calculated by using the  short wave approximation method. We conclude the paper in Section (V) with brief comments.  As far as we know, gravitational waves in the BD$\Lambda$ theory has not been studied and our contribution in this work is novel.

\section{Weak Field Equations}

 The Einstein- Hilbert (EH) action with a cosmological constant is given by,
\begin{equation} \label{action1} 
S_{GR\Lambda } = \int \sqrt{\left|g\right|} d^{4}  x\left[\frac{1}{2\kappa } (R-2\Lambda )+L_{m} \right] 
\end{equation} 
where $\kappa ^{2} =\frac{8\pi G}{c^{4} } $ is the  gravitational coupling constant, R is the Ricci scalar, $\left|g\right|$ is the absolute value of the determinant  of the metric tensor $g_{\mu \nu } $ and $\Lambda $ is the cosmological constant term. We may call the theory implied by the action (\ref{action1}), as General Relativity  theory in the presence of a cosmological constant  (GR$\Lambda$). Throughout this study we use the units in which $c=G=1$ and we use a metric signature  $(-, +, +, +)$. 

Scalar tensor theories are some of the most studied alternative theories of gravity  and  BD theory of gravity is the simplest one of those theories providing a very suitable test bed of the prediction of theories alternative to general Relativity. This theory includes both a scalar field  $\phi$ and metric tensor $g_{\mu \nu }$ to describe the gravitational interaction. Hence, gravitational interactions are partly due to the curvature of the space-time and are partly due to the effect of the scalar field. In the original Brans-Dicke theory, cosmological constant was not included, hence  $\Lambda $ is equal to zero in original BD action. But, in  this paper  we will consider  BD theory with a cosmological constant. To derive such a theory from GR$\Lambda$ theory with  EH action extended with a cosmological constant, we can replace the gravitational coupling constant  $\kappa$,  with  a scalar field, namely we can set $\kappa \to 8\pi \phi ^{-1} $ in the action (\ref{action1}) as done in the original BD theory. It is clear that the BD scalar field is inversely proportional to the gravitational coupling constant. A very  straightforward extension of GR$\Lambda$ theory to BD scalar-tensor theory is just following the original BD prescription by replacing Newton coupling constant $G$ with a scalar field $\phi$ and adding a dynamical term coupled by an arbitrary parameter $\omega$ which is called as the BD parameter. Hence the  action of this theory can be described by the following action \cite{Uehara} in Jordan frame, in which the matter Lagrangian is not coupled to the scalar field, as: 
\begin{equation} \label{BDLambdaaction} 
S_{BD\Lambda } =\int \sqrt{\left|g\right|}  d^{4} x\left\{\frac{1}{16\pi } \left[\phi (R-2\Lambda )-\frac{\omega }{\phi } g^{\mu \nu } \partial _{\mu } \phi\, \partial _{\nu } \phi \right]+L_{m} \left[\psi _{m} ,g_{\mu \nu } \right]\right\}, 
\end{equation} 
where $\phi $ is BD scalar field and $\omega $ is the dimensionless BD parameter. As we have said before, this action may be called as the BD theory with a cosmological constant (BD$\Lambda $). When setting $\Lambda=0$, this theory reduces to the original form of the BD theory \cite{BransDicke}. As $\phi $ becomes constant, this theory reduces to GR$\Lambda $ theory. But this action is not the only one which reduces to GR$\Lambda $ when $\phi $ becomes constant. If we replace $2\Lambda \phi $ term with an arbitrary potential term $V(\phi )$
in (\ref{BDLambdaaction}) we obtain an action which corresponds to a theory which may be  called as  the BD theory with a potential (BDV theory), where $V(\phi )$ acts as a variable cosmological term \cite{Ozer}.
 If $\phi $ is set to a  constant, this action also reduces to GR action with a cosmological constant. We may call this latter theory as BDV theory.  Clearly, there is an arbitrariness in the generalization of GR with a cosmological constant to BD theories. One may ask why consider the theories as different theories since $2\Lambda \phi$ term in (\ref{BDLambdaaction}) can be included as a special case of $V(\phi)$. 
 The answer is yes, because BD$\Lambda$ and BDV theories have several important different properties in which some of them is given below:
\begin{itemize}
	\item BD$\Lambda$ theory has a massless scalar field whereas  in BDV theory  the scalar field attains an  effective mass.
	\item In BD$\Lambda$ theory scalar field has a long range similar to the original BD theory, keeping that property of BD theory while introducing a cosmological constant term. Whereas in BDV theory the mass of the scalar field makes the scalar field a short range one.  
\end{itemize}

We believe that  due to the differences summarized above,  these theories deserve a separate analysis  in order to investigate possible  different physical consequences  of these theories. However since gravitational waves of BD$V$ theory or massive BD theory, which also includes $f(R)$ theory as a special case, is already investigated in great detail in the flat background case, we only  consider gravitational wave solutions  of BD$\Lambda$ theory and refer the works \cite{Alsing,Maggiore,Will} for corresponding solutions in BDV theory.

The field equations of the action (\ref{BDLambdaaction}) can be expressed as
\begin{eqnarray} \label{BDFE}
	&&G_{\mu\nu}+\Lambda\, g_{\mu\nu} =\frac{8\pi}{\phi}T_{\mu\nu}+\frac{\omega}{\phi^2}\left(\nabla_\mu \phi \nabla_\nu \phi -\frac{1}{2}g_{\mu\nu} \nabla^\alpha \phi \nabla_\alpha\phi \right)+\frac{1}{\phi} \left(\nabla_\mu \nabla_\nu\phi-g_{\mu\nu} \Box_g \phi \right)
	,\\
	&& \Box_g\phi=\frac{1}{2\omega+3}\left(8\pi\, T-2 \Lambda\,\phi  \right),
\end{eqnarray}
where  $T=T^\mu_{\phantom{a}\mu}$  is the trace of the energy-momentum tensor $T_{\mu\nu}$ and $\Box_g$ is the D'Alembert operator with respect to the metric $g_{\mu\nu}$.   In order to obtain the weak field expansion of the above field equations, we can expand  the space time metric and the BD scalar field as
\begin{eqnarray}\label{weak1}
&&g_{\mu\nu}=\eta_{\mu\nu}+h_{\mu\nu},\quad g^{\mu\nu}=\eta^{\mu\nu}-h^{\mu\nu},\\
&&\phi=\phi_0+\varphi, \nonumber
\end{eqnarray}  
where $\eta_{\mu\nu}=\mbox{diag}(-1,1,1,1)$ is the Minkowski metric, $h_{\mu\nu}$  is the metric perturbation tensor representing small deviation from flatness, $\phi_0$ is a constant value of the scalar field and $\varphi$ is a small perturbation to the scalar field i.e., $|h_{\mu\nu}| \ll 1$  and  $\varphi \ll 1$. The linearised field equations can be expressed in a very economical way using the above expansion, by defining a new tensor \cite{Will}
\begin{equation}\label{weak2}
\theta_{\mu\nu}=h_{\mu\nu}-\frac{1}{2} \eta_{\mu\nu}h-\eta_{\mu\nu} \frac{\varphi}{\phi_0},
\end{equation}
 and considering  the gauge
\begin{equation}\label{gauge-cond}
\theta^{\mu\nu}_{\phantom{aa};\nu}=0.
\end{equation}
Finally, the weak BD  field equations can be expressed, up to \emph{second order}, as
 \begin{eqnarray}\label{weak-second-order}
&&\Box_\eta\theta_{\mu\nu}=-\frac{16 \pi}{\phi_0}\left(T_{\mu\nu}+\tau_{\mu\nu} \right)+2\Lambda \eta_{\mu\nu}+\frac{2\Lambda \varphi}{\phi_0}\eta_{\mu\nu},
\\
&&\Box_\eta \varphi=16 \pi S. \label{weak-second-order-sc}
\end{eqnarray} 
Here $\tau_{\mu\nu}$ is the energy-momentum pseudo tensor involving quadratic terms and $\Box_\eta=\eta^{\mu\nu}\partial_\mu \partial_\nu$ is the D'Alembert operator of the Minkowski spacetime. The corresponding expressions in the presence of an arbitrary potential can be found in \cite{Will}. The term $S$ is given by 
\begin{equation}\label{S}
	S=\frac{1}{4\omega+6} \left[ T \left(1-\frac{\theta}{2}-\frac{\varphi}{\phi_0} \right)-\frac{\Lambda}{4\pi}\left( \phi_0+\varphi \right)  \ \right]+\frac{1}{16\pi}\left( \theta^{\mu\nu}\varphi_{,\mu\nu}+\frac{\varphi_{,\nu}\varphi^{,\nu}}{\phi_0}\right).
\end{equation}
To obtain the expression of $S$,  the relation between Minkowski and curved D'Alembert operators is used \cite{Will}:
\begin{equation}
\Box_g=\left(1+\frac{\theta}{2}+\frac{\varphi}{\phi_0} \right)\Box_\eta-\theta^{\mu\nu}\varphi_{,\mu\nu}-\frac{\varphi_{,\nu}\varphi^{,\nu}}{\phi_0}+O(\mbox{higher order terms}).
\end{equation}

In this study one of the our aim is to obtain the linearised field equations of BD theory in the presence of  a constant background curvature coupled to the scalar field in a straightforward way. Now we will discuss these cases respectively.

\subsection{Linearised Field Equations of Brans-Dicke Theory with a Cosmological Constant 
}

 We will consider the weak field expanded equations of the  action \eqref{BDLambdaaction}, given in equations (\ref{weak-second-order},\ref{weak-second-order-sc}), in the linear order in the parameters $\Lambda, \varphi, h_{\mu\nu} (\mbox{or } \theta_{\mu\nu})$   by ignoring $\Lambda \varphi $ terms or other second and higher order terms.  Then the linearised field equations become,
\begin{equation} \label{BDLambdaFE1} 
\square_\eta {\rm }\theta _{\mu \nu } =-\frac{16\pi }{\phi _{0} } T_{\mu \nu } +2\Lambda \eta _{\mu \nu },                                                                                                 
 \end{equation}                                                                                                                             \begin{equation}\label{BDLambdaFE2}
 \square_\eta {\rm }\varphi =\frac{8\pi T}{{\rm (2}\omega {\rm +3)}} -\frac{2\Lambda \phi _{0} }{{\rm (2}\omega {\rm +3)}}.  
\end{equation} 
As seen from the above equations, tensor equation has similar structure to GR$\Lambda$ theory \cite{Bernabeu} but there is also a scalar field equation unlike GR$\Lambda $ theory and the scalar field is massless as in BD theory. Here the scalar field has a long range, the cosmological constant plays the role of a background curvature as in GR$\Lambda $ theory and the existence of the cosmological constant does not change this property.

\subsection{ Point  Mass Solutions of  linearised Field Equations}

 We  have presented the linearised field equations under weak field expansion and in the certain  gauge for  BD$\Lambda $ theory.  Before delving into gravitational wave solutions, let us review localized point mass solution of this theory. The point mass solution, in the presence of cosmological constant or for non vanishing minimum  potential, was presented and  their physical properties  were  discussed comparatively in detail   in \cite{Ozer}. Hence in the following we just give the result in the isotropic spherical coordinates for BD$\Lambda$ theory.

For a point particle  at the origin having  the energy momentum tensor as $ T_{\mu \nu } =m\, \delta (r)\, \mbox{diag}(1,0,0,0)$, the weak field equations (\ref{BDLambdaFE1},\ref{BDLambdaFE2})  can be solved in Cartesian coordinates respecting the Loretz gauge and can be transformed into  isotropic spherical coordinates as explained in \cite{Ozer}. The result is

\begin{equation} \label{GrindEQ__32_} 
g_{00} =-1+\frac{2m}{\phi _{0} r'} \left(1+\frac{1}{2\omega +3} \right)+\frac{\Lambda r'^{2} }{3} \left(1-\frac{1}{2\omega +3} \right),                                                                
\end{equation}
\begin{equation}\label{GrindEQ__33_} 
 g_{ij} =\delta _{ij} \left[1+\frac{2m}{\phi _{0} r'} \left(1-\frac{1}{2\omega +3} \right)-\frac{\Lambda r'^{2} }{6} \left(1-\frac{2}{2\omega +3} \right)\right],
\end{equation}
\begin{equation}\label{GrindEQ__34_} 
\phi =\phi _{0} \left(1+\frac{2m}{(2\omega +3)\phi _{0} r'} -\frac{\Lambda r'^{2} }{3(2\omega +3)} \right). 
\end{equation} 
The mass term in $g_{00} $ must be identical to weak field  GR or Newton potential of a point mass. For this reason  $\phi _{0} $ must be equal to,
\begin{equation} \label{GrindEQ__35_} 
\phi _{0} =\frac{2\omega +4}{2\omega +3}  
\end{equation} 
Also in the absence of  $\Lambda $, all of the metric and field components reduce to the point mass solution of linearised BD theory \cite{BransDicke,Brans}. Moreover  these solutions also reduce to linearised GR$\Lambda $ solutions in the limit of ($\omega \to \infty $,$\phi _{0} \to 1$ ) \cite{Bernabeu}.  Thus,  the solutions \eqref{GrindEQ__32_} and \eqref{GrindEQ__33_} have the correct limits.
This solution clearly shows  that  in  the BD$\Lambda$ theory the scalar field has long range, unlike BD$V$ theory where mass term involves a Yukawa type  term making the field a short range one \cite{Will,Alsing}.

\section{Gravitational Waves in Brans-Dicke Theory with a cosmological constant}

In the previous section, we have investigated  BD theories whose weak field equations are of the form \eqref{BDLambdaFE1} and \eqref{BDLambdaFE2}. Now we want to obtain the wave solutions of these theories, respectively. Let us expand the space time metric as,
\begin{equation} \label{GrindEQ__58_} 
	g_{\mu \nu } =\eta _{\mu \nu } +h^{\Lambda } {}_{\mu \nu } +h^{W} {}_{\mu \nu },                                                                                                    
\end{equation}
\begin{equation}
	h^{\Lambda ,W} {}_{\mu \nu }  \ll 1 \label{GrindEQ__59_}, 
\end{equation} 
where  $h_{\mu \nu }^{\Lambda } $ is the background perturbation due to cosmological term and $h_{\mu \nu }^{W} $ is the gravitational wave perturbation. In the same approach the tensor $\theta _{\mu \nu } $  and the scalar field $\varphi$ can be expanded as,
\begin{equation} \label{GrindEQ__60_} 
	\theta _{\mu \nu } =\theta ^{\Lambda } {}_{\mu \nu } +\theta ^{W} {}_{\mu \nu },  
\end{equation}
and 
\begin{equation} \label{GrindEQ__61_} 
	\varphi =\varphi ^{\Lambda } +\varphi ^{W}.  
\end{equation} 
As it is clear from Eq. \eqref{GrindEQ__58_}, \eqref{GrindEQ__60_} and \eqref{GrindEQ__61_}, both background modification and gravitational wave perturbation affects the field equations. Hence we must consider both of these contributions. Firstly, we calculate the effects of the background perturbation on the linearised field equations by ignoring possible ripples in the spacetime.

\subsection{Background solutions}

 For background solutions, linearised field  equations \eqref{BDLambdaFE1}, \eqref{BDLambdaFE2} and the gauge condition \eqref{gauge-cond} take the following form, 
\begin{equation} \label{GrindEQ__62_} 
\square{\rm }\theta ^{\Lambda } {}_{\mu \nu } =2\Lambda \eta _{\mu \nu },  
\end{equation} 
\begin{equation} \label{GrindEQ__63_} 
\square\varphi ^{\Lambda } =-\frac{2\Lambda \phi _{0} }{2\omega +3} , 
\end{equation} 
\begin{equation} \label{GrindEQ__64_} 
\theta ^{\Lambda ^{} \mu } {}_{\nu ,\mu } =0. 
\end{equation} 
The solutions of the field equations, after imposing the Lorentz gauge, become,
\begin{equation} \label{GrindEQ__65_} 
\theta ^{\Lambda } {}_{\mu \nu } =-\frac{\Lambda }{9} x_{\mu } x_{\nu } +\frac{5\Lambda }{18} \eta _{\mu \nu } x^{2},                                                                                             
\end{equation}

\begin{equation} \label{GrindEQ__67_} 
\varphi ^{\Lambda } =-\frac{\Lambda \phi _{0} }{3(2\omega +3)} r^{2}.  
\end{equation} 
We have chosen the background scalar solution $\varphi^\Lambda$ as proportional to $r^2$ in (\ref{GrindEQ__67_}) rather than $x^2=\eta_{\mu\nu}x^\mu x^\nu$ as  $\varphi ^{\Lambda } =-\frac{\Lambda \phi _{0} }{4(2\omega +3)} x^{2}$. This choice is made due to the fact that in the linearised theory we want the background solution to be static as in line with the Newtonian theory. The tensor part can be a function of time, since one can remove that dependence by a suitable coordinate transformation as in the GR case \cite{Bernabeu} but we feel that for the scalar field we need to implement this staticity condition by hand. Note that unlike GR there is no Birkhoff theory for BD. Hence there is a possibility that weak field time dependent solutions are  also possible in spherical Schwarzschild coordinates in dS backgrounds. But we will not pursue this case in this paper.  

By reverting the equation (\ref{weak2}), we can obtain the metric perturbation tensor as follows 

\begin{equation} \label{GrindEQ__66_}                                    
h^{\Lambda } {}_{\mu \nu } =-\frac{\Lambda }{9} x_{\mu } x_{\nu } -\frac{2\Lambda }{9} \eta _{\mu \nu } x^{2} -\eta _{\mu \nu } \frac{\varphi ^{\Lambda } }{\phi _{0} }.  
\end{equation} 
In the case of vanishing scalar field, this result reduces exactly to the result of \cite{Bernabeu}.

Using these solutions, the space-time metric becomes,
\begin{eqnarray}
ds^{2} =&&-\left[1+\frac{\Lambda }{9} (3t^{2} -2r^{2} )-\frac{\varphi ^{\Lambda } }{\phi _{0} } \right]dt^{2} +\left[1-\frac{\Lambda }{9} (-2t^{2} +2r^{2} +x^{i2} )-\frac{\varphi ^{\Lambda } }{\phi _{0} } \right]dx^{i2}\nonumber
\\
&&+\frac{2\Lambda }{9}\, t\,x^{i} dt \,dx^{i} -\frac{2\Lambda }{9}\, x^{i}\, x^{j}\, dx^{i}\, dx^{j},  
\label{GrindEQ__68_}                                          
\end{eqnarray}
\noindent where  $i=1,2,3$ and $i\ne j$. This line element, although it respects the gauge condition we are considering, is neither homogeneous nor isotropic. In order to compare this metric with the observations, it might be useful to transform this solution in a better known forms such as a homogeneous and isotropic form, with the help of  appropriate coordinate transformations. Firstly we apply a similar coordinate transformations to convert the solution into the static solution as done in GR$\Lambda$ case \cite{Bernabeu1,BernabeuE}:
\begin{equation}
 \begin{array}{l} {x=\tilde{x}+\frac{\Lambda }{9} \left(-\tilde{t}^{2} -\frac{\tilde{x}^{2} }{2} +(\frac{\tilde{y}^{2} +\tilde{z}^{2} }{4} )\right)\tilde{x}} \\ 
{y=\tilde{y}+\frac{\Lambda }{9} \left(-\tilde{t}^{2} -\frac{\tilde{y}^{2} }{2} +(\frac{\tilde{x}^{2} +\tilde{z}^{2} }{4} )\right)\tilde{y}} \\ 
{z=\tilde{z}+\frac{\Lambda }{9} \left(-\tilde{t}^{2} -\frac{\tilde{z}^{2} }{2} +(\frac{x'^{2} +\tilde{y}^{2} }{4} )\right)\tilde{z}}                                                                           \\
{t=\tilde{t}-\frac{\Lambda }{18} (\tilde{t}^{2} +\tilde{r}^{2} )\tilde{t}}\label{GrindEQ__69_} \\
\end{array} 
\end{equation}
where $\tilde r^{2} =\tilde{x}^{2} +\tilde{y}^{2} +\tilde{z}^{2}. $ 
When we keep only first order terms, the resulting  metric becomes,

\begin{equation} \label{GrindEQ__70_} 
ds^{2} =-\left[1-\frac{\Lambda }{3} \tilde{r}^{2} -\frac{\tilde{\varphi }^{\Lambda } }{\phi _{0} } \right]d\tilde{t}^{2} +\left[1-\frac{\Lambda }{6} (\tilde{r}^{2} +3\tilde{x}_{i} {}^{2} )-\frac{\tilde{\varphi }^{\Lambda } }{\phi _{0} } \right]d\tilde{x}_{i} {}^{2}  
\end{equation} 
\begin{equation} \label{GrindEQ__71_} 
\tilde{\varphi}^\Lambda=-\frac{\Lambda \phi _{0} }{3(2\omega +3)} \tilde{r}^{2}.  
\end{equation} 
We can obtain a spherically symmetric solution, when we apply the same change of coordinates presented in \cite{Bernabeu},
\begin{equation} \label{GrindEQ__72_} 
\begin{array}{l} {\tilde{x}=x'+\frac{\Lambda }{12} x'^{3} }, \\ {\tilde{y}=y'+\frac{\Lambda }{12} y'^{3} }, \\ {\tilde{z}=z'+\frac{\Lambda }{12} z'^{3} }, \\ {\tilde{t}=t'}. \end{array} 
\end{equation} 
Under these transformations, the metric and the scalar field takes the form,
\begin{equation} \label{GrindEQ__73_} 
ds^{2} =-\left[1-\frac{\Lambda }{3} r'^{2} -\frac{\varphi '^{\Lambda } }{\phi _{0} } \right]dt'^{2} +\left[1-\frac{\Lambda }{6} r'^{2} -\frac{\varphi '^{\Lambda } }{\phi _{0} } \right](dr'^{2} +r'^{2} d\Omega ^{2} ), 
\end{equation} 
\begin{equation} \label{GrindEQ__74_} 
\varphi'^{\Lambda}=-\frac{\Lambda \phi _{0} }{3(2\omega +3)} r'^{2}.  
\end{equation} 

We have converted the solution into the homogeneous and isotropic form.
This background solution has the same result with the point mass solution of BD$\Lambda $ theory presented in \cite{Ozer} given in equations (\ref{GrindEQ__32_})--(\ref{GrindEQ__34_}), if the mass is equal to zero.

 Let us apply another coordinate transformation to make this metric similar to the Schwarzschild-de Sitter (SdS ) metric,               
\begin{equation} \label{GrindEQ__75_} 
\begin{array}{l} { r'=r\left(1+\frac{\Lambda }{12} r^{2} +\frac{\varphi ^{\Lambda } }{\phi _{0} } \right)}, \\ {t'=t}. \end{array} 
\end{equation} 

This transformation brings the line element (\ref{GrindEQ__73_})  to the following expression:  
\begin{equation} \label{GrindEQ__76_} 
ds^{2} =-\left[1-\frac{\Lambda }{3} r^{2} -\frac{\varphi ^{\Lambda } }{\phi _{0} } \right]dt^{2} +\left[1+\frac{\Lambda }{3} r^{2} +\frac{\alpha r}{\phi _{0} } \right]dr^{2} +r^{2} d\Omega ^{2} , 
\end{equation} 
where we have made  a new definition,
\begin{equation} \label{GrindEQ__77_} 
\alpha =\frac{d\varphi }{dr} =-\frac{2\Lambda \phi _{0} r}{3(2\omega +3)}.  
\end{equation} 
The scalar field has the following form 
\begin{equation} \label{GrindEQ__74_} 
\varphi^{\Lambda} =-\frac{\Lambda \phi _{0} }{3(2\omega +3)} r^{2}.  
\end{equation} 
 This Schwarzschild type form of the linearised background solution may be useful for future applications.
\subsection{Wave solutions of BD$\Lambda$ theory}

 Now let us consider the second case  where  a  gravitational wave  solution exists by taking  into account the gravitational wave perturbations that may occur under the presence of a source which may cause fluctuations in the space-time geometry and the scalar field. For this case the gauge condition can be written as,
\begin{equation} \label{GrindEQ__78_} 
\theta ^{W\mu \nu } {}_{,\mu } =0.
\end{equation} 

Then the total wave equation is given by,
\begin{equation} \label{GrindEQ__80_} 
{\rm \square}\left(\theta ^{\Lambda } {}_{\mu \nu } +{\rm }\theta ^{W} {}_{\mu \nu }\right) =2\Lambda \eta _{\mu \nu }.  
\end{equation} 
The homogeneous part of the field equation \eqref{GrindEQ__80_}   is sufficient to represent the gravitational  waves. Hence we can write the wave equation as,
\begin{equation} \label{GrindEQ__81_} 
\square{\rm }\theta ^{W} {}_{\mu \nu } =0 
\end{equation} 
Similarly for the scalar field,
\begin{equation} \label{GrindEQ__82_} 
\square\varphi=\square(\varphi^\Lambda+\varphi^W) = -\frac{2\Lambda\phi_0}{2\omega+3}
\end{equation} 
the homogeneous part 
\begin{equation}\label{GrindEQ__82_} 
\square \varphi^W =0, 
\end{equation}
will give the wave solution.

The solution of the field Eq.  \eqref{GrindEQ__81_} can be written as, 
\begin{equation} \label{GrindEQ__83_} 
\theta ^{W} {}_{\mu \nu } =A_{\mu \nu } \sin kx+B_{\mu \nu } \cos kx 
\end{equation} 
where $A_{\mu \nu } $ and $B_{\mu \nu } $ are amplitude tensors and $\mathbf{k}=(k^0,\vec{k})$ is a wave four-vector, i.e., $kx=k_{\mu } x^{\mu } =k_{0} t+\vec{k}.\vec{x}$. 
 If we plug the solution \eqref{GrindEQ__83_} into the Eq. \eqref{GrindEQ__78_} and \eqref{GrindEQ__81_}, we find,
\begin{eqnarray} \label{GrindEQ__84_} 
&&k_{\mu } A {}^{\mu } {}_{\nu } =k_{\mu } B^{\mu } {}_{\nu } =0,\\ 
&&k^{2} =k_{\mu}k^{\mu}=0, \label{GrindEQ__85_}\\
&&{A^W}_{\mu }^{\mu } ={B^W}_{\mu }^{\mu } =0.\label{transverse}
\end{eqnarray} 
Eq. \eqref{GrindEQ__84_} shows that the amplitude tensors  $A_{\mu \nu } $ and $B_{\mu \nu } $ are orthogonal to the direction of the propagation of the wave and Eq. \eqref{GrindEQ__85_} shows the fact that the gravitational waves propagates  at the speed of light.   Hence, considering also  (\ref{transverse})  we can conclude that the tensorial part of the  gravitational waves in BD$\Lambda$ theory  are transverse and traceless waves moving with the speed of light. 

 The solution (\ref{GrindEQ__83_})  describes a wave with the angular frequency,
\begin{equation} \label{GrindEQ__86_} 
\nu =k^{0} =(k_{x} {}^{2} +k_{y} {}^{2} +k_{z} {}^{2} )^{{1\mathord{\left/ {\vphantom {1 2}} \right. \kern-\nulldelimiterspace} 2} }.
\end{equation}

The solution of the scalar field equation \eqref{GrindEQ__82_} is given by, 
\begin{equation} \label{GrindEQ__87_} 
\varphi ^{W} =C\sin kx+D\cos kx, 
\end{equation} 
where C and  D are integration constants. Using Eq. \eqref{GrindEQ__83_} and \eqref{weak2}, the total metric perturbation induced by some source of GW can be written as
\begin{equation}
 h_{\mu \nu }^{W} =A_{\mu \nu } \sin kx+B_{\mu \nu } \cos kx-\frac{\eta _{\mu \nu } }{\phi _{0} } (C\sin kx+D\cos kx).                                                                 \label{GrindEQ__88_}
\end{equation}

The total solution of linearised field equations of BD theory involving a cosmological constant can be written  for the tensor $\theta_{\mu\nu}$ as
\begin{equation} \label{GrindEQ__92_} 
\theta _{\mu \nu } =\theta ^{\Lambda } {}_{\mu \nu } +\theta ^{W} {}_{\mu \nu } =-\frac{\Lambda }{9} x_{\mu } x_{\nu } +\frac{5\Lambda }{18} \eta _{\mu \nu } x^{2} +A_{\mu \nu } \sin kx+B_{\mu \nu } \cos kx 
\end{equation} 
and finally the metric perturbation tensor becomes
\begin{eqnarray}
 h_{\mu \nu } &=&h^{\Lambda } {}_{\mu \nu } +h^{W} {}_{\mu \nu } =-\frac{\Lambda }{9} x_{\mu } x_{\nu } - \left(\frac{2\Lambda x^2 }{9}+\frac{\varphi^\Lambda}{\phi_0} \right)\eta _{\mu \nu }  \nonumber \\
&& +A_{\mu \nu } \sin kx+B_{\mu \nu } \cos kx-\frac{\eta _{\mu \nu } }{\phi _{0} } (C\sin kx+D\cos kx)
  \label{GrindEQ__93_}
\end{eqnarray} 
where the scalar field $\varphi^{\Lambda}$ is given in (\ref{GrindEQ__67_}).

The  total scalar field solution in a constant curvature background is given by
\begin{equation} \label{GrindEQ__94_} 
\varphi =\varphi ^{\Lambda } +\varphi ^{W} =-\frac{\Lambda \phi _{0} }{3(2\omega +3)} r^{2} +C\sin kx+D\cos kx. 
\end{equation}

\subsection{The Effects of GW on Free Particles in BD$\Lambda$ theory}

Here, we want to understand the effect of the gravitational waves in BD$\Lambda $ theory on free particles or detectors.  Since the weak equivalence principle holds for BD theory a single particle cannot feel the metric perturbations.  The simplest way to understand the physical effects of GW on matter  is to consider the relative motion of two nearby test particles in free fall.  Hence, let us consider two nearby freely falling particles of equal mass and specify the spacetime coordinates  of these particles with $(t,x,y,z)$ \cite{Inverno,Maggiore}.  Then, we calculate the geodesic deviations caused by the GW. The geodesic deviation equation is defined as \cite{Padmanabhan}
\begin{equation} \label{GrindEQ__129_} 
\frac{d^{2} \zeta ^{\alpha } }{d\tau ^{2} } =R_{\phantom{a}\beta \nu \chi }^{\alpha } U^{\beta } U^{\nu } \zeta ^{\chi },  
\end{equation} 
where $\zeta ^{\mu } $ is the  vector  connecting the particles , $U^{\mu } $ is the four velocity of the  two particles and in the rest frame of the observer \cite{Rindler},
\begin{equation} \label{GrindEQ__130_} 
U^{\mu } =\left(1,0,0,0\right), 
\end{equation} 
and  we  assume a wave moving along the z direction, then the wave vector becomes,
\begin{equation} \label{GrindEQ__131_} 
k^{\mu } =(\nu ,0,0,\nu ). 
\end{equation} 
Inserting Eq. \eqref{GrindEQ__130_} into the Eq. \eqref{GrindEQ__129_},  we obtain,
\begin{equation} \label{GrindEQ__132_} 
\frac{d^{2} \zeta ^{\alpha } }{d\tau ^{2} } =- R_{\phantom{a}0i0}^{\alpha } \zeta^i.  
\end{equation} 
This result shows that the Riemann tensor is locally measurable by calculating the separation between nearby geodesics.

 For the calculational simplicity, here we consider a gravitational wave moving along $z$ direction.  We transform the trigonometric functions  such that  $A_{\mu\nu} \sin(kx)+B_{\mu\nu} \cos{kx}=\tilde A_{\mu\nu} \cos(kx-\delta) $ where $\delta$ is phase of the wave and we also set $\delta=0$ for simplicity. We also use that for waves moving on $z$ direction $kx=\nu(t-z)$. 
Then the general solution can be simply written as
 \begin{eqnarray} \label{GrindEQ__133_} 
 h_{\mu \nu }&=&h_{\mu\nu}^{\Lambda}+h_{\mu\nu}^W =-\frac{\Lambda }{9} x_{\mu } x_{\nu } -\frac{2\Lambda }{9} \eta _{\mu \nu } x^{2} -\eta _{\mu \nu } \frac{\varphi ^{\Lambda } }{\phi _{0} } +\left( \tilde{A}_{\mu \nu }-\eta_{\mu\nu}\tilde D  \right) \cos \nu(t-z)  \\
 	\varphi &=&\varphi ^{\Lambda } +\varphi ^{W} =-\frac{\Lambda \phi _{0} }{3(2\omega +3)} r^{2} +D\cos \nu (t-z). 
 \end{eqnarray}
where $\tilde A_{\mu\nu} $ is the amplitude tensor adapted to the problem by taking real part of the full solution with an appropriately chosen phase and $\tilde D=D/\phi_0$ is the real part of the scalar amplitude constructed similarly.

Since Riemann tensor is gauge invariant, we can use the linear form of Riemann tensor in TT gauge \cite{Adler} to calculate the nonvanishing components of this tensor  given by
\[
2 R_{abcd}=
h_{ad,bc}-h_{bd,ac}+h_{bc,ad}-h_{ac,bd}
\]

Using this  linearised form of Riemann tensor and the solution (\ref{GrindEQ__133_}), the necessary components  are found as

\begin{eqnarray}
R^x_{\phantom{a}0x0}&=&\frac{1}{2} \left[\left(\tilde A_{xx}-\tilde D \right)\nu^2-\frac{2\Lambda }{3}\left(1-\frac{1}{2\omega+3} \right)  \right],\\
R^y_{\phantom{a}0x0}&=&R^{x}_{\phantom{a}0y0}=\frac{1}{2} \nu ^{2} \tilde  A_{xy},\\
R^y_{\phantom{a}0y0}&=& \frac{1}{2} \left[\left(-\tilde A_{xx}-\tilde D \right)\nu^2 -\frac{2\Lambda }{3}\left(1-\frac{1}{2\omega+3} \right) \right],\\
R^z_{\phantom{a}0z0}&=&\frac{1}{2} \left(-\frac{2\Lambda }{3}\left(1-\frac{1}{2\omega+3} \right)  \right)\\
R^z_{\phantom{a}0x0}&=&R^z_{\phantom{a}0y0}=0.
\end{eqnarray}

Now, suppose that the first particle is at the origin and second one is at the point $\zeta^i=(\zeta,0,0)$,  so that the separation between the particles is $\zeta $ and these particles are initially at rest. We will analyze how this seperation changes in the presence of an incident gravitational wave propagating in the $z$ direction.
In this case, the relative acceleration of freely falling test particles become,
\begin{eqnarray} \label{GrindEQ__134_} 
\frac{d^{2} \zeta ^{x} }{d\tau ^{2} } &=&-\zeta R_{\phantom{a}0x0}^{x}=\frac{\zeta}{2} \left[-\left(\tilde A_{xx}+\tilde D \right)\nu^2+\frac{2\Lambda }{3} \left(1-\frac{1}{2\omega+3} \right) \right],\\                                                                                                                    \frac{d^{2} \zeta ^{y} }{d\tau ^{2} } &=&-\zeta R_{\phantom{a}0x0}^{y}= -\frac{\zeta}{2} \nu ^{2} \tilde  A_{xy},\label{GrindEQ__135_} \\                                                                                                                  \frac{d^{2} \zeta ^{z} }{d\tau ^{2} } &=&-\zeta R_{\phantom{a}0x0}^{z}=0. \label{GrindEQ__136_}  
\end{eqnarray}      

 Similarly, the relative acceleration of two nearby freely falling particles seperated by $\zeta $ in the $y$ direction become,
 \begin{eqnarray} \label{GrindEQ__144_} 
 &&\ddot{\zeta }^{x} =-\frac{\zeta }{2} \nu ^{2} \tilde A_{xy} \\                                                                                                               &&\ddot{\zeta }^{y} =\frac{\zeta }{2} \left[(\tilde A_{xx}-\tilde D)\nu ^{2} +\frac{2\Lambda }{3} \left(1-\frac{1}{2\omega+3} \right) \right] \label{GrindEQ__145_} \\                                                    
 && \ddot{\zeta }^{z} =\frac{d^{2} \zeta ^{z} }{d\tau ^{2} } =0.\label{GrindEQ__146_} 
 \end{eqnarray} 
For the third case, lets consider two nearby particles initially seperated by $\zeta $ in the $z$ direction. The  relative acceleration obey,
\begin{eqnarray} \label{GrindEQ__147_} 
&&\ddot{\zeta }^{x} =0,  \\
 \label{GrindEQ__148_} 
&&\ddot{\zeta }^{y} =0,\\
 \label{GrindEQ__149_}  
&&\ddot{\zeta }^{z} =\frac{\zeta }{2} \left[\frac{2\Lambda}{3} \left(1-\frac{1}{2\omega+3} \right)\right]. 
\end{eqnarray} 
The terms proportional to $\Lambda$ mimick usual homogeneous expansion of the universe due to the cosmological term. The terms proportional to $\tilde A_{xx}$ and $\tilde A_{xy}$ denotes the effects of the usual plus and cross polarizations of the gravitational wave and the terms proportional to $\tilde D$ corresponds to the scalar breathing mode of the Brans-Dicke theory.
So a gravitational wave detector feels extra stretching between arms due to the cosmological constant which is practically zero for a detector in the surface of earth due to the smallness of the cosmological constant. The effect of the BD field is to reduce the amount of this stretching compared to GR and  for the critical value $\omega_{c\Lambda}=-1$ the effect of the cosmological constant is balanced by the attractive effect of the BD scalar. If $\omega<-1$ the combined effect becomes negative implying an attractive effect despite we have an asymptotically dS spacetime.  In summary,  we  have a negligible pressure  between arms of the detector due to the combined effect of cosmological constant and BD field which is equal to the same pressure in GR due to positive cosmological constant as $\omega\rightarrow \infty$.  This effect gets smaller by decreasing $\omega$  which vanishes for $\omega=\omega_{c\Lambda}$  and becomes a tension with values  of the BD parameter smaller that $\omega_{c\Lambda}$.
Considering all these, we see that,  if the wave is propagating along the $z$ direction, the metric perturbation can be expressed as a sum of three polarization states,  
\begin{equation}
 h_{\mu \nu }^{\Lambda} (t-z)=A^{+} (t-z)e_{\mu \nu }^{+} +A^{\times } (t-z)e_{\mu \nu }^{\times } +\Phi (t-z)\eta _{\mu \nu }                                                    \label{GrindEQ__150_} 
\end{equation}
 where $e_{\mu \nu }^{+} $ and $e_{\mu \nu }^{\times } $ denote the usual plus and cross polarization tensors, respectively of gravitational waves, and  $A^{+} $, $A^{\times } $ are amplitudes of these tensors \cite{Rindler}:
\begin{equation} \label{GrindEQ__151_} 
e_{\mu \nu }^{+} =\left(\begin{array}{cccc} {0} & {0} & {0} & {0} \\ {0} & {1} & {0} & {0} \\ {0} & {0} & {-1} & {0} \\ {0} & {0} & {0} & {0} \end{array}\right),\quad  e_{\mu \nu }^{\times } =\left(\begin{array}{cccc} {0} & {0} & {0} & {0} \\ {0} & {0} & {1} & {0} \\ {0} & {1} & {0} & {0} \\ {0} & {0} & {0} & {0} \end{array}\right),\quad     \eta _{\mu \nu } =\left(\begin{array}{cccc} {-1} & {0} & {0} & {0} \\ {0} & {1} & {0} & {0} \\ {0} & {0} & {1} & {0} \\ {0} & {0} & {0} & {1} \end{array}\right). 
\end{equation} 
In other words, in BD$\Lambda $ theory, there are two massless spin 2 modes and  one massless scalar mode \cite{Will,Alsing,Brunetti,Maggiore}. In the  geodesic deviation equations,  $\tilde A_{xx} $ term denotes the plus mode  and $\tilde A_{xy} $ term denotes the cross mode. The terms involving the cosmological constant $\Lambda$  give the contribution of the cosmological constant on the polarization states. Also, the  terms proportional to $\tilde D$   denote the breathing mode resulting from the massless scalar field.  Therefore,  we can conclude that the BD$\Lambda $ theory  has three polarization states, two of which denote plus and cross polarization states and the other one denotes breathing mode caused by the massless scalar field. This is what we were expecting. This result is in agreement  with that obtained in original BD theory without a cosmological constant or a potential and as we have seen from the analysis of BD$\Lambda$ theory in \cite{Ozer}, the cosmological term does not attain a mass to the scalar field in local gravitational sectors. As we have shown explicitly, this property persists for gravitational waves as well.   Besides, the effect of the existence of the cosmological constant on the geodesic deviation equations is the same amount of cosmological acceleration in all directions, mimicking the homogeneous cosmological expansion due to cosmological constant.

In order to see these results in an other perspective, we transform the wave solutions into FRW coordinates. Hence we may also apply the following change of coordinates  \cite{Bernabeu1,BernabeuE},
\begin{equation} \label{GrindEQ__90_} 
	\begin{array}{l} {x^{i} =e^{T\sqrt{{\Lambda \mathord{\left/ {\vphantom {\Lambda  3}} \right. \kern-\nulldelimiterspace} 3} } } X^{i} }, 
		\\ {t=\frac{1}{2} \sqrt{\frac{\Lambda }{3} } R^{2} +T}, \end{array}                                 
\end{equation} 
where X,Y,Z are comoving coordinates and  $R=\sqrt{X^{2} +Y^{2} +Z^{2} } $ . As a result of the calculations, the transformed wave-like solution to order $\sqrt{\Lambda } $ becomes,
\begin{eqnarray}
	\label{GrindEQ__91_} 
	h_{\mu\nu}^W &=& \left[ \begin{array}{cccc} {\tilde D } & {0} & {0} & {0} \\ {0} & (\tilde A_{xx} -\tilde D  ) \left(1+2\sqrt{\frac{\Lambda }{3} } T \right) & {\tilde A_{xy} \left(1+2\sqrt{\frac{\Lambda }{3} } T\right)} & {0} \\ {0} & {\tilde A_{xy} \left(1+2\sqrt{\frac{\Lambda }{3} } T \right)} & {(-\tilde A_{xx} -\tilde D )\left(1+2\sqrt{\frac{\Lambda }{3} } T \right)} & {0} \\ {0} & {0} & {0} & {-\tilde D \left(1+2\sqrt{\frac{\Lambda }{3} } T \right)} \end{array} \right]\nonumber 
	\\
	\times &&\cos \left(\nu (T-Z)+\nu \sqrt{\frac{\Lambda }{3} } \left(\frac{Z^{2} }{2} -TZ \right)+O(\Lambda )\right) 
\end{eqnarray}

These results clearly shows the effects of the GR wave on the spacetime. The three polarization degrees of freedom will have same dispersion relation
\begin{equation}\label{dispersion}
	\nu (T-Z)+\nu\sqrt{\frac{\Lambda}{3}}\left( \frac{Z^2}{2}-TZ \right)=n\,\pi,
\end{equation}
which yields approximately
\begin{equation}\label{Zmax}
	Z_{max}(n,T)\simeq T-\frac{n\pi}{\nu}-\sqrt{\frac{\Lambda}{3}}\left(\frac{T^2}{2}-\frac{n^2\,\pi^2}{2\,\nu^2} \right)
\end{equation}
In these expressions the Brans-Dicke parameter does not enter in the equations. Hence, these expressions showed that the behaviour of tensor part of gravitational waves behave exactly the same as in GR$\Lambda$ theory \cite{Bernabeu} regarding the frequency of the waves. The scalar waves, which does not exists in GR, also shows the same behaviour with the tensor ones. 

In order to obtain the linearised solutions in FRW type coordinates which can represent the universe we live in, we have first solved the linearised equations in the coordinates and the gauge where the linearised equations are valid. The transformed solutions do not satisfy the linearised BD$\Lambda$ field equations. The important point is the first order correction in FRW type coordinates in the cosmological constant is in the order of $\sqrt{\Lambda}$ rather than $\Lambda$  as in the coordinates respecting linearised equations. Hence, for the observational point of view the effects of $\Lambda$ is much more relevant despite the smallness of the observed value of cosmological constant.

\subsection{Comparison With Observations and  Other Theories}
In this part we will comment on some properties of gravitational waves of BD$\Lambda$ theory  in comparison with other modified gravity theories, especially with BDV theory and $f(R)$ theories, since these properties may be important to distinguish this theory with other  alternative gravity theories. However due to the enormous number of modified gravity theories exist in the literature, we are not  attempting to list  or  compare   these theories one by one  with BD$\Lambda$ theory, but   
we prefer to refer some of the renowned reviews \cite{Sotiriou,Nojiri,Capozziello,Clifton, Koyama,Will} for a list and properties and possible tests of  these theories. Here we consider the two important aspects of gravitational waves,  polarization properties and propagation speeds of  these modes, that may differ for different gravitation theories \cite{Will}, the properties  which are independent from the source of the waves.  

 In a generic gravity theory, a gravitational wave can have six independent polarization modes in which two of them are  tensorial, two of them are scalar and  the remaining two  are vectorial. Among these modes, two tensorial modes, namely the $+$ and $\times$ modes and also the scalar breathing mode are transverse, and remaining longitudinal scalar mode and the two vectorial modes  oscillate in the direction on propagation. See \cite{Will} for a detailed description  of these modes with figures.  
Although the gravitational waves in general relativity have only the well known  $+$ and $\times $  tensorial modes, a particular alternative theory may have, in addition to usual tensorial modes, some of the other four extra modes as well. For example,  in scalar-tensor theories including BD theory,  one or two scalar modes can be present, depending on whether the scalar is massless or massive \cite{Will}.  Moreover, vector-tensor theories can have additional vector modes \cite{Will}  and scalar-vector-tensor theories can have all types of polarization states, for example there exists five polarization modes in Einstein-Aether theories \cite{Jacobson} whereas  six polarization modes in TeVeS \cite{Teves1}.
  If we can observe polarization modes 
  of waves from a given source, we can determine the  viability of a given modified gravity theory. Unfortunately, due to their orientation, the two LIGO detectors are not capable to detect extra modes, and the addition of Virgo detector did not improve this fact much yet. Therefore, the polarization properties of waves is not yet a viability criteria for modified gravity theories. The situation is hoped to improve in the future, among other ways, when new earth based dedectors will  start to make observations producing an effective network of dedectors, which is believed to be sensitive to detect these extra polarization modes \cite{Chatziioannou:2012rf}. 

Another important property of gravitational waves are the propagation speeds of their independent  modes.  In GR, both modes move with the speed of light, however in some of the alternative theories some or all modes may have speeds different than speed of light.  Hence the speed of waves is another property distinguishing different gravitation theories regarding observational data if we can observe the speed of propagation of waves from a given source.   For the speed of waves we are more lucky because in addition to the gravitational waves, we can also detect possible electromagnetic signals from the sources if they emit. In this regard, the discovery of merger of two neutron stars \cite{Neutron}  turn out to be very important. The reason for this is that, unlike black hole merger events, electromagnetic waves are also detected \cite{gamma} in this event.  Detecting signals from  various channels of these kind of sources  opens a new and great way to test general relativity and its alternatives. One can determine the speed of gravitational waves by comparing the arrival times of gravitational and electromagnetic waves from these sources to our detectors. 
Actually, in this observation, both gravitational \cite{Neutron} and electromagnetic \cite{gamma} waves coming from  the merger of two neutron stars producing a BH as an end point was observed nearly simultaneously. This observation opened a new era called multi-messenger astronomy, which has important consequences on testing modified theories using gravitational and electromagnetic wave observations. 
The observations \cite{Neutron,gamma} showed that the gravitational waves (or at least $+$ and $\times$ modes) move with the speed of light since, as a result of these observations, the difference between speed of gravitational waves, $c_{gw}$, and speed of light, $c$, is limited by the inequality $3\times 10^{-15}<|(c_{gw}-c)/c|<7\times 10^{-16}$. This also implies that some of the alternative theories proposing  smaller speeds for tensorial modes of these waves are no longer viable \cite{creminelli,sakstein,ezquiaga,baker} such as Covariant Galileons, Fab Four, Gauss-Bonnet and some subcases of beyond Horndeski theory if they also claim to explain the dynamics of the universe.   Note that some other theories, including  the BD theory and its extensions discussed in this paper and $f(R)$ theories are among the surviving theories \cite{casalino} since they pass the observational constraints imposed by this observation. We refer to the latest reviews \cite{Ezquiagarev,BertiRev} about the status of modified theories on this observation. 

Let us discuss in more detail the  polarization modes of BD$\Lambda$ theory.  Similar to the  BD theory and unlike massive BD ( or BDV) theory, we have found in this paper that the scalar gravitational waves in BD$\Lambda$ theory are transverse, massless and  move with the speed of light. The tensorial modes are the same with those of GR, so we focus on the   third polarization mode, the scalar breathing mode, present in BD$\Lambda$ theory, which is the main difference between GR and BD theories regarding the gravitational waves. What we found in  Section (III.C) is that, similar to BD theory, there is no longitudinal scalar mode in BD$\Lambda$ theory. In comparison, the scalar  part of the gravitational waves in the BDV theory, however, is massive, moving slower than speed of light and  has in addition a longitudinal mode where the motion takes place in the direction of propagation of the wave \cite{Alsing,Maggiore,Will}. These are the differences between gravitational waves in BD($\Lambda$) and BDV theories. Now let us compare the results we have obtained with the  gravitational waves  in $f(R)$ theories \cite{CapoziellofR,Corda,Berry,BerryE, Rizwana,Shaoi,Dicong}, since these theories also have scalar modes. It is known that  BDV theory is equivalent to $f(R)$ theories for particular values of BD parameter $\omega$ such as for $\omega=0$ metric and $\omega=-3/2$ Palatini $f(R)$ theories \cite{Sotiriou,Clifton,Capozziello}.  Hence the distinction for BD$\Lambda$ and BDV theories also persist for $f(R)$ theories, despite the character of the extra modes stem from the scalar field in BDV theory and the curvature scalar in $f(R)$ theories.   Namely, unlike BD$\Lambda$ theory, the scalar mode in $f(R)$ theory is massive, moves slower than speed of light, and have both breathing  and  longitudinal modes. 
Note that these differences cannot be detected in current dedectors, since the dedectors under operation can only feel the $+$ and $\times$ modes, rendering BD, BD$\Lambda$, BDV and $f(R)$ theories compatible with recent observations regarding both polarization properties and speed of scalar waves. However, when the number of new detectors making observations increase in the near future, with the possibility of detecting  scalar modes and their propagation speeds, we may hope to investigate the viability of these scalar tensor theories.  Especially, restrictions on the parameter spaces of these kind of theories might emerge in the future by using the observational data of a possible network of multiple dedectors,  by the addition of new dedectors to LIGO and Virgo, since such a  network  may  be capable  \cite{Chatziioannou:2012rf} of detecting scalar and other modes.



\section{Energy-Momentum tensor of Gravitational Waves }

In this section we calculate the gravitational energy carried by the gravitational waves in BD$\Lambda$ theory as well as the energy of the background geometry. To calculate the background energy due to cosmological constant in BD theory we first expand metric tensor as $g_{\mu\nu}=g_{\mu\nu}^{(b)}+h_{\mu\nu}$ where $g_{\mu\nu}^{(b)}$ is a background metric. The scalar field  is also expanded as $\phi=\phi_0+\varphi$ as given in (\ref{weak1}). Using these, then we express  relevant scalars and tensors in terms of them as  in the expansion of the Ricci tensor is $R_{\mu\nu}=R_{\mu\nu}^{(b)} +R_{\mu\nu}^{(1)}+R_{\mu\nu}^{(2)}\ +\ldots$. Using these we can express the field equations in the orders of $O(h,\varphi)$. 
We use the following expansion of the Ricci tensor in the first and second order \cite{MTW}, where the bar denotes covariant differentiation, in our calculations:
\begin{eqnarray}
	R_{\mu \nu }^{(1)}&=&\frac{1}{2}\left(-h_{|\mu\nu}-h_{\mu\nu|\alpha}^{\phantom{aaia}\alpha} +h_{\alpha\mu|\nu}^{\phantom{aaai} \alpha}+ +h_{\alpha\nu|\mu}^{\phantom{aaai} \alpha}\right),\\ 
	R_{\mu \nu }^{(2)}&=&\frac{1}{2}\left[\frac{1}{2} h_{\alpha\beta|\mu} h^{\alpha\beta}_{\phantom{aa}|\nu} +h^{\alpha\beta}\left( h_{\alpha\beta|\mu\nu}+ h_{\mu\nu|\alpha\beta}-h_{\alpha\mu|\nu\beta}-h_{\alpha\nu|\mu\beta}\right)
	\right. \\ &&\left.
	+ h_{\nu}^{\phantom{a}\alpha|\beta}\left( h_{\alpha\mu|\beta} - h_{\beta\mu|\alpha}  \right)  -\left(h^{\alpha\beta}_{\phantom{aa}|\beta}-\frac{1}{2}h^{|\alpha} \right)\left( h_{\alpha\mu|\nu}+h_{\alpha\nu|\mu}-h_{\mu\nu|\alpha} \right) \right].             
\end{eqnarray}

We also  consider the fact that the linearized equations are satisfied for first order terms in the field equations and the second order terms contribute to the energy of the background geometry. Therefore the energy momentum tensor due to cosmological constant in BD theory can be calculated by subtracting first order terms from total expressions keeping only the second order terms. We choose background geometry as the flat Minkowski spacetime. The energy momentum tensor of $\Lambda$, hence becomes:
\begin{eqnarray}\label{tmunubackground}
t_{\mu\nu}=\frac{\phi_0}{8\pi}\left( R_{\mu\nu}^{(2)}-\frac{1}{2}\eta_{\mu\nu}R^{(2)}+\frac{1}{2}\eta_{\mu\nu}h^{\alpha\beta}R_{\alpha\beta}^{(1)} -\Lambda  h_{\mu\nu}+T_{\mu\nu}[\varphi^2]\right)
\end{eqnarray}  
with the contribution of the BD scalar to the energy-momentum tensor of the waves is identified as
\begin{equation}\label{EMPhi}
	T_{\mu\nu}(\phi)=\frac{\omega}{\phi^2}\left(\phi_{,\mu}\phi_{,\nu}-\frac{1}{2}g_{\mu\nu}\, \phi^{,\alpha} \phi_{,\alpha} \right)+\frac{1}{\phi}\left(\phi_{,\mu\nu}-g_{\mu\nu}\Box  \phi \right)
\end{equation}
where $T_{\mu\nu}[\varphi^2]$ in (\ref{tmunubackground}) represents the second order contributions due to the scalar and metric perturbations $\varphi$ and $h_{\mu\nu}$ to $T_{\mu\nu}(\phi)$  given in (\ref{EMPhi}). 

After a long calculation of all relevant tensorial components of the background solution (\ref{GrindEQ__67_},\ref{GrindEQ__66_}) we have found the Poynting vector due to cosmological constant as
\begin{equation}\label{PointingLambda}
	t_{0z}^\Lambda=\frac{\phi_0}{8\pi}\left[\frac{22}{81}-\frac{4}{9(2\omega+3)} \right]\Lambda^2\, t\,z
\end{equation}
where the effect of the scalar field is negative for $\omega>-2/3$ and positive for $\omega<-2/3$.

The energy-momentum carried by the waves in BD theory is well known \cite{Will,Alsing}. Here we derive them using the  method known as the short wave approximation method  \cite{Brill,Isaacson1,Isaacson,MTW}. This method considers a general gravitational wave in a curved vacuum background geometry. There are two relevant length scales in this method:  The first one, $L$,  represents the typical curvature radius and the second one, $\lambda$, represents a typical wavelength of the waves with the assumption that $\lambda\ll L$. 
Essence of this method considers again the splitting of the metric of the spacetime into a  background metric  which is a slowly changing function of spacetime and also the metric  perturbation $h_{\mu\nu}$ representing high frequency waves  with $\alpha$ being the amplitude of waves. Now the perturbation metric satisfies $h_{\mu \nu } \leqslant g_{\mu \nu }^{(B)} \times \alpha$.  In this approximation the derivatives of the metric components vary as $g_{\mu \nu,\alpha }^{(B)} \leqslant \frac{g_{\mu \nu } }{L}, $ and $ h_{\mu \nu,\alpha} \leqslant \frac{h_{\mu \nu } }{\lambda }$. Using steady coordinates, we find the typical order of magnitudes of Ricci tensors as $R_{\mu \nu }^{(B)} \sim \frac{\alpha }{\lambda ^{2} }$, $R_{\mu \nu }^{(1)} \sim \frac{\alpha }{\lambda ^{2} }$, $R_{\mu \nu }^{(2)} \sim \frac{\alpha ^{2} }{\lambda ^{2} }$. For our calculations we choose background as  Minkowski spacetime, and also split $h_{\mu\nu}=h_{\mu\nu}^\Lambda +h_{\mu\nu}^W$, signifying the cosmological and wave contributions to the metric perturbation tensor. We can also split the scalar field as $\phi=\phi^{(b)}+\varphi$, where $\varphi$ may contain wave part as well as other perturbative corrections to the chosen background scalar $\phi^{(b)}$. In this paper we choose $\phi^{b}=\phi_0$ and $\varphi=\varphi^\Lambda+\varphi^W$ where former denote the cosmological background and latter denote the wave contribution to the scalar field.  

The short wave approximation requires solving the vacuum BD$\Lambda$ equations under the  circumstances summarized in the previous paragraph. This implies that the field equations  already satisfy the linearized  equations. Hence the contribution of the waves to the energy-momentum tensor shows themselves in the second order. Second order field equations contain parts due to cosmological background as well as due to the fluctuations of the spacetime and the  scalar field.  For example, the Ricci tensor can be written as parts containing background free of ripples as well as parts containing the wave part as follows: The smooth part has the terms: $R_{\mu\nu}=R_{\mu\nu}^{(b)}+<R_{\mu\nu}^{(2)}>+\,\mbox{error}$ and the fluctuating part contains: $R^{(1)}(h^2)+R^{(2)}_{\mu\nu}(h)-<R_{\mu\nu}^{(2)}(h)>+\,\mbox{error}$. Note that in GR, both expressions equal to zero due to the vacuum equations require $R_{\mu\nu}=0$, separately. However, in BD there are also scalar contributions to the field equations that we will considered in the following.  The operation  $<...>$  symbolizes averaging of quantity, which is required since the linearized tensors that we are considering  such as Ricci tensor and also  the energy-momentum tensor are not gauge invariant in these linearized forms. By averaging over several wavelengths,  we  may hope to include enough curvature in a small region to make these quantities gauge invariant \cite{Brill,Isaacson,Isaacson1,MTW}.  We can identify, as done in \cite{Brill,MTW,Isaacson,Isaacson1} in GR, the background Einstein tensor in vacuum proportional to the energy-momentum tensor of the waves, meaning that the energy-momentum of the waves generates the background curvature. Hence, we obtain that
\begin{equation}
	G_{\mu\nu}^{(b)} \equiv R_{\mu\nu}^{(b)}-\frac{1}{2}\,g_{\mu\nu}^{(b)}R^{(b)}=8\pi\, T_{\mu\nu}^{(W)}
\end{equation}
where  the energy momentum tensor of the  gravitational waves is given by \begin{equation}\label{emtgw}
	T_{\mu\nu}^{W}=\frac{\phi_0}{8\pi}\left\{-<R_{\mu\nu}^{(2)} (h)>+\frac{1}{2}g_{\mu\nu}^{(b)}<R^{(2)}(h)>+\frac{1}{2}h_{\mu\nu}R^{(1)}-\Lambda <h_{\mu\nu}>+<T_{\mu\nu}^{(2)}(\varphi^2)> \right\}
\end{equation}
where again  $T_{\mu\nu}(\varphi^2)$ represents the second order contributions of the BD scalar due to wave parts of the perturbations into energy-momentum tensor of gravitational waves identified in (\ref{EMPhi})   and evaluated at the second order in $h_{\mu\nu}$ and $\varphi$.

Using the  tensor $\theta_{\mu\nu}$ defined in $(\ref{weak2})$ along with gauge (\ref{gauge-cond}) simplifies the energy-momentum tensor given in (\ref{emtgw}). Moreover, we use  an  important advantage of using averaging process  that the averages of first derivatives vanish as $<\partial_\mu X>=0$, which also implies that $<A\,\partial_\mu B>=-<(\partial_\mu A )\,B>$.   After a long calculation  using all these, we found the result
\begin{equation}\label{energywave}
	t^W_{\mu\nu}=\frac{\phi _{0} }{32\pi } \left\{\left\langle \theta _{\alpha \beta \left|\mu \right. } \theta ^{\alpha \beta } {}_{\left|\nu \right. } \right\rangle +\frac{(4\omega +6)}{\phi _{0} {}^{2} } \left\langle \varphi _{\left|\mu \right. } \varphi _{\left|\nu \right. } \right\rangle  - 4\left\langle \Lambda h_{\mu \nu } \right\rangle \right\},
\end{equation}
which agrees with previous results \cite{Will,Alsing,Brunetti} derived using other methods. Evaluating this formula for the solution that we have derived in (\ref{GrindEQ__133_}) corresponding to gravitational waves moving in $z$ direction, and calculating the averages gives the following result for the Poynting vector 
\begin{equation} \label{pointing} 
	t_{03}^{W} =\frac{\phi _{0} }{32\pi } \left[-\nu ^{2} \left\{ \left|A_{+} \right|^{2} +\left|A_{\times } \right|^{2} +\frac{(2\omega +3)}{\phi _{0} {}^{2} } |D|^{2} \right\} \right] 
\end{equation}
 where we have only used wave parts of the solutions to avoid double counting since the pointing vector for cosmological constant is already calculated in (\ref{PointingLambda}).

The total Poynting vector combining contributions of cosmological constant and GR waves becomes
\begin{equation} \label{GrindEQ__188_} 
	t_{03} =\frac{\phi _{0} }{32\pi } \left\{\left(\frac{88 }{81}-\frac{16}{9(2\omega +3)} \right)\Lambda ^{2}\, t\,z-\nu ^{2} \left[ \left|A_{+} \right|^{2} +\left|A_{\times } \right|^{2} +\frac{(2\omega +3)}{\phi _{0} {}^{2} } |D|^{2} \right] \right\}
\end{equation}
where $\nu $ is the frequency of the wave and $A_{+} $ and $A_{\times } $ are plus and cross polarization tensors, respectively. The first two terms depend on the cosmological constant and shows the contribution of the cosmological constant to the energy flux. The third and fourth terms are the energy flux of GR theory \cite{MTW,Brill1} and the last term is the contribution of the massless  BD scalar field to the energy flux \cite{Will,Alsing, Brunetti}. 

 Now we can define a critical distance  as in \cite{Arraut,Nowakowski} as the distance where 	$t_{03}$ vanishes, given by $t_{03}^\Lambda=t^W_{03}$. Since both scalar and gravitational waves move with speed of light in BD$\Lambda$ theory, we can identify the time with a distance in (\ref{PointingLambda}).  In terms of observable frequency $f=\nu/2\pi$, and curvature radius of dS spacetime $L=\sqrt{3/\Lambda}$, the critical distance  become 
\begin{equation}\label{Lc}
	L_c=\frac{2\pi f A}{\sqrt{a} \Lambda},
\end{equation}
where
 \begin{equation}
a=\frac{88 }{81}-\frac{16}{9(2\omega +3)}, \quad A=\sqrt{\left|A_{+} \right|^{2} +\left|A_{\times } \right|^{2} +\frac{(2\omega _{BD} +3)}{\phi _{0} {}^{2} } |D|^{2} }.
 \end{equation}

The expression we have derived is similar to what obtained in \cite{Nowakowski,Arraut} with some differences in numerical factors.  The critical distance  result is valid in the region $\omega<-3/2$ or $\omega>-15/22$ due to $\sqrt{a}$ term. The effect of the cosmological constant is to hinder the propagation of gravitational waves  distances beyond this length. The waves cannot propagate beyond $L_c$, since space asymptotically goes to dS spacetime where the inhomogeneities such as gravitational waves vanish. As discussed in \cite{Arraut}, this is due to cosmic no-hair conjecture (CNC) \cite{Bicak}, which requires that the inhomogeneities to be dissipated at large distances and large times. Since $r_{\Lambda}=\sqrt{3/\Lambda}$ and $f=1/\lambda \approx r_\Lambda$ \cite{MTW}, we see that the critical distance is  proportional to dS radius and hence the background scale of the spacetime , $L_c  \sim r_\Lambda$.  This proves that the  CNC  is also valid for gravitational waves in BD theory in the presence of a positive cosmological constant where inhomogeneities cannot be propagated further from dS radius of the spacetime.
 We have, in this paper, extended this result to the BD theory in the presence of the cosmological constant $\Lambda$. A confirmation of this result can be done as a future work by studying the global structure of  the Robinson-Trautman type radiative spacetime solutions in BD theory \cite{Tahamtan,Dereli}, as done in \cite{Bicak2}.

\section{CONCLUSION}

\noindent In this study  weak field gravitational wave solutions of  Brans-Dicke theory with a cosmological constant (BD$\Lambda $) is obtained in the presence of a positive cosmological constant. It is known that the scalar field is massless for this theory and hence the range of the scalar field is still a long range one, a behaviour contrary to masssive BD theory where the existence of an arbitrary potential introduces an effective mass and makes the scalar field short range one. Hence, in this theory, the effect of the cosmological constant is to behave like a constant background curvature. Therefore, in this paper, what we have found are the weak field wave solutions on the spacetime having a constant curvature background in BD theory.

After obtaining these solutions we have discussed their physical properties. We have seen that these waves behave similarly to gravitational waves in the original BD  theory, rather than massive BD or $f(R)$ theories. The weak field  gravitational effects  of wave like perturbations on constant positive background curvature of BD theory consist of three parts. We have two tensorial waves having usual plus and cross polarizations. These waves are  traceless and transverse waves  moving with speed of light. The third component, the scalar wave,  is massless and moves also with speed of light.

We have studied the effects of these waves on test particles and dedectors by calculating  the geodesic deviation by the gravitational wave of two test particles initially at rest. We see that, we have usual plus and cross polarizations as in GR and in addition there is a scalar breathing mode as in the BD theory.
 We see that the combined  effect of the cosmological constant and the background scalar field  is to exert an homogeneous expanding force on these particles to extend the length between them similar to Hubble expansion of two galaxies due to  the dark energy. Note that the force due to scalar background is attractive and acts in the opposite direction which increases with decreasing BD parameter, $\omega$, such that at $\omega=-1$ these two effects cancel each other with no net force between two particles due to $\Lambda$. When $\omega<-1$, although we have a positive cosmological constant,  the force between these test particles become attractive.

 We have also calculated  the energy-momentum tensor of gravitational waves for the BD$\Lambda $ theory calculated by using the shortwave approximation method. The obtained results give the energy-momentum tensor  due to the  background curvature and background scalar field as well as those of gravitational waves of this theory. Although the energy momentum tensor of gravitational waves in  the original and massive BD theories is well known, we have obtained those in the BD$\Lambda$ theory using the shortwave approximation method which is  a different method than the method used  to calculate the previous results in original and massive BD theories in for example \cite{Will}. The energy of the background fields is new in this theory. Analyzing the energy flux for waves propagating in a particular direction, we see that gravitational waves can not be propagated distances further  than the  background scale, namely that $L_c \approx r_\Lambda$, given in  (\ref{Lc}). This results extend the  Cosmic No hair Theorem (CNC) to BD theory in the presence of a cosmological constant for  the region $\omega<-3/2$ or $\omega>-15/22$.

 The solutions and analysis obtained within the scope of this study can be used to test the predictions of BD$\Lambda$ theory with the observational data accumulated by the LIGO and Virgo observatories.  Current wave observations are consistent with the predictions of the general relativity which claim that gravitational waves propagate at the speed of light and posses two polarization states. Note that the interferometers used in these observatories are just sensitive to plus and cross gravitational waves and hence cannot test the  scalar breathing modes. Hence,  BD  and BD$\Lambda$ theories are not limited or ruled out  by the current data. To test scalar modes, we need either a spherical dedector or a network of interferometers distributed on the surface of earth.   If the scalar mode would be  massive, then the differences of  arrival times of waves from the same event  between  scalar and tensorial waves to LIGO and VIRGO or other dedectors on the different points of earth could be used to test  that theory and limit its parameters such as mass of the scalar field. However,  this is not the case for BD$\Lambda$ theory since the scalar waves  are massless and move with the speed of light. Extending the results of the observations of LIGO and Virgo and future possible  observations of  other experiments currently in different stages of operation to  test the massless or massive breathing or longitudional waves of these type of theories might be a possible next  direction of the research of gravitational waves in alternative theories of GR.

\begin{acknowledgements}
	H. O. is partially supported by the TUBITAK 2211/E Programme. This paper is partially based on H.O.'s PhD thesis \cite{Hatice}.
\end{acknowledgements}


\begin{thebibliography}{widestlabel}
	\bibitem{MTW} C. W. Misner, K. S. Thorne and J. A. Wheeler, Gravitation, (Freeman, NY, 1974).
		\bibitem{Sotiriou} T. P. Sotiriou  and V. Faraoni,  Rev. Mod. Phys. 82, 451 (2010).
	\bibitem{Nojiri} 
	S. Nojiri and S. D. Odintsov,
	Phys. Rept. 505, 59 (2011).
 \bibitem{Capozziello} S.Capozziello and M.De Laurentis, Phys. Rept. 509, 167 (2011).
	\bibitem{Clifton} T. Clifton, P. G. Ferreira, A. Padilla and C. Skordis, Phys. Rep. 513, 1 (2012).
\bibitem{Utiyama} R. Utiyama ans B. S.  DeWitt, J. Math. Phys. 3, 608 (1962).
\bibitem{Stelle} K. S. Stelle, Phys. Rev. D 16, 953 (1977).  
\bibitem{Lidsey} J. E. Lidsey, D. Wands and E. Copeland, Phys. Rep. 337, 343 (2000).
\bibitem{Kaluza} T. Kaluza, Sits. Preus. Akad. Wiss. 33, 966 (1921).
\bibitem{Klein} O. Klein, Z. Phys, 37, 895 (1926).

	\bibitem{Einstein}  A. Einstein, Sitzungsberichte der Preussischen Akademie der Wissenschaften zu Berlin, 22, { 688} (1916).
	
	\bibitem{indirect } R.  A.  Hulse  and  J.  H.  Taylor,  Astrophys.  J.  Lett., 195,  L51 (1975).
	
	\bibitem{detection} B. P. Abbott et al. (LIGO Scientific Colloboration, Virgo Colloboration) Phys. Rev. Lett. 116, 061102 (2016). 
	\bibitem{properties}  B. P. Abbott et al. (LIGO Scientific Colloboration, Virgo Colloboration) Phys Rev. Lett. 116, 241102 (2016). 
	
	\bibitem{Neutron} B. P. Abbott et al. (LIGO Scientific Colloboration, Virgo Colloboration)  Phys. Rev. Lett. 119, 161101 (2017).  
  \bibitem{GWTC-1} B. P. Abbot. et al. (LIGO Scientific Collaboration and Virgo Collaboration) Phys. Rev. X 9,031040 (2019).
  \bibitem{GWTC-2} B. P. Abbot. et al. (LIGO Scientific Collaboration and Virgo Collaboration) arXiv:2010.14527 [gr-qc]
	\bibitem{test} B. P. Abbott et al. (LIGO Scientific Colloboration, Virgo Colloboration)   Phys. Rev. Lett. 116, 221101 (2016).
\bibitem{test1}  B. P. Abbott et al. (LIGO Scientific Colloboration, Virgo Colloboration)  Phys. Rev. D 100, 104036 (2019).
	\bibitem{test2}    B. P. Abbott et al. (LIGO Scientific Colloboration, Virgo Colloboration) arXiv:2010.14529 [gr-qc]
	
		\bibitem{Will} Will C.M., Theory and Experimentation in Gravitational Physics, Second  Edition, (Cambridge University Press, Cambridge, U.K, 2018)
	\bibitem{BransDicke} C. Brans and R. H. Dicke, Phys. Rev. 124, 925 (1961).
	\bibitem{Brans} C. H. Brans, Phys. Rev. 125, 2194 (1962).
	
	\bibitem{Bertotti} B. Bertotti, L. Iess and P. Tortora, Nature 425, 374 (2003).
	
	\bibitem{Pimentel} L. O. Pimentel, Astrophys. Space Sci. 112, 175 (1985).
		\bibitem{Kolithic} S. J.  Kolitch, Annals Phys. 246, 121 (1996).

\bibitem{Romero1} C. Romero and A. Barros,  Astrophys. Space Sci. 192, 263 (1992). 	
		\bibitem{Romero} C. Romero and A. Barros,  Gen. Relativ. Gravit.  25, 491 (1993).
		\bibitem{Tretyakova:2011ch} 
	D.~A.~Tretyakova, A.~A.~Shatskiy, I.~D.~Novikov, and S.~Alexeyev,
	Phys.\ Rev.\ D  85, 124059 (2012).
		\bibitem{delice1}O. Delice,  Phys. Rev. D 74, 067703 (2006). 
		\bibitem{Aguledo} J. A. Aguledo, J. R. Nascimento, A. Yu Petrov, P. J. Porfirino, and A. F. Santos Phys: Lett. B 762, 96 (2016).  	
	
		\bibitem{Oh} J. Lee, T. H. Lee, and P. Oh, Phys. Lett. B, 701 393 (2011) . 

	
		\bibitem{Barrow} J. D. Barrow and K. Maeda, Nucl. Phys. B  341, 294 (1990).
	
		\bibitem{Uehara} K. Uehara and C. W. Kim, Phys. Rev. D  26, 2575 (1982).
	\bibitem{Lorenz} D. Lorenz-Petzold, Phys. Rev. D 29, 2399 (1984).
	\bibitem{Lorenz1} D. Lorenz-Petzold, Prog. Theor. Phys.  71, 1426 (1984).
	
	
	
	
	\bibitem{Pandey} S. N. Pandey, Astrophys. Space Sci. 277, 403 (2001).
	
	\bibitem{Novikov}I. D. Novikov, A. A. Shatskii, S. O. Alexeyev, and D. A. Tret'yakova, Phys. Usp.  57, 352 (2014). 
	
	
	\bibitem{delice2}O. Delice,  Phys. Rev. D  74, 124001  (2006). 
	\bibitem{baykaldeliceciftci} A. Baykal, D. K. Ciftci, and O. Delice,  J. Math. Phys.  51, 072505 (2010).	
		\bibitem{Ozer} H. Ozer and O. Delice,  Class. Quantum Grav. 35, 065002 (2018).
	
	
		\bibitem{Alsing} J. Alsing, E. Berti, C. M. Will, and H. Zaglauer, Phys. Rev. D 85, 064041 (2012).
\bibitem{Maggiore} M. Maggiore and A. Nicolis,  Phys. Rev. D 62, 024004, (2000).


	\bibitem{Bernabeu} J. Bernabeu, C. Espinoza, and N. E. Mavromatos, Phys. Rev. D 81, 084002 (2010).

	 
	\bibitem{Bernabeu1}J. Bernabeu, D. Espriu and D. Puigdomenech,  Phys. Rev. D 84, 063523 (2011). 
		\bibitem{BernabeuE}	J. Bernabeu, D. Espriu and D. Puigdomenech,  Phys. Rev. D 86, 069904(E) (2012).
	
	\bibitem{Inverno} R.  D'Inverno, Introducing Einstein's Relativity, (Oxford Univ. Press, Oxford UK, 1990) 

	\bibitem{Padmanabhan}  Padmanabhan, T., Gravitation (Foundations and Frontiers), (Cambridge University Press, Cambridge UK 2010)
	
		\bibitem{Adler} R. Adler, M. Bazin and M. Schiffer, Introduction to General Relativity, (Mcgraw-Hill press, Second Edition, 1975)

	\bibitem{Rindler} W. Rindler, Relativity, Special, General and Cosmological, 2nd.  Edition, (Oxford University Press, Oxford UK, 2006)
	
		\bibitem{Brunetti} M. Brunetti, E. Coccia, V. Fafone, and F. Fucito, Phys. Rev. D 59, 044027 (1999).
	
\bibitem{Koyama} K.~Koyama,
Rept. Prog. Phys. 79, 046902 (2016).


\bibitem{Jacobson}
T.~Jacobson and D.~Mattingly,
Phys. Rev. D 70, 024003 (2004).
	
	\bibitem{Teves1} 
	J.~D.~Bekenstein,
		Phys. Rev. D 70, 083509 (2004).
	
	\bibitem{Chatziioannou:2012rf}
	K. Chatziioannou, N. Yunes and N.Cornish,
	Phys. Rev. D 86, 022004 (2012).
	
	
	\bibitem{gamma}  B. P. Abbott et al., (Virgo, Fermi-GBM, INTEGRAL, LIGO Scientific collaboration), Astrophys. J. 848, L13 (2017).
	\bibitem{creminelli} P. Creminelli and F. Vernizzi, Phys. Rev. Lett. 119, 251302 (2017).
	\bibitem{sakstein} J. Sakstein and B. Jain, Phys. Rev. Lett. 119, 251303  (2017).
	\bibitem{ezquiaga} J. M. Ezquiaga and M. Zumalacárregui,Phys. Rev. Lett. 119, 251304  (2017). 
	\bibitem{baker} T. Baker, E. Bellini, P. G. Ferreira, M. Lagos, J. Noller and I. Sawicki, Phys.Rev. Lett. 119, 251301   (2017). 

\bibitem{casalino} 
A.~Casalino, M.~Rinaldi, L.~Sebastiani and S.~Vagnozzi,
Class. Quant. Grav. 36, 017001 (2019).

\bibitem{Ezquiagarev}
J. M. Ezquiaga and M. Zumalac\'arregui,
Front. Astron. Space Sci. 5, 44 (2018).
\bibitem{BertiRev} 
E.~Berti, K.~Yagi and N.~Yunes,
Gen. Relativ. Gravit. 50,  46 (2018).
	\bibitem{CapoziellofR} 
S. Capozziello, C. Corda and M. F. De Laurentis,
Phys. Lett. B 669, 255 (2008).
\bibitem{Corda} C. Corda, Eur. Phys. J. C 65  257 (2010).
\bibitem{Berry} C. P. L. Berry and J. R. Gair, Phys. Rev. D 83, 104022 (2011). 
\bibitem{BerryE}C. P. L. Berry and J. R. Gair, Phys. Rev. D 85, 089906(E) (2012). 
\bibitem{Rizwana}
H. Rizwana Kausar, L. Philippoz and P. Jetzer,
Phys. Rev. D 93, 124071 (2016).

\bibitem{Dicong} D. Liang, Y. Gong, S. Hou and Y. Liu, Phys.  Rev.  D 95, 104034 (2017).
\bibitem{Shaoi} S. Hou, Y. Gong and Y. Liu,  Eur. Phys. J. C  78, 378 (2018). 


		\bibitem{Brill}  D. R. Brill, J. B. Hartle,  Phys. Rev. 135, B271 (1964).
	
	\bibitem{Isaacson} R. A. Isaacson,  Phys. Rev. 166, 1263 (1968).
	
	\bibitem{Isaacson1}  R. A. Isaacson,  Phys.  Rev. 166, 1272 (1968).
	\bibitem{Brill1}  B. F. Schutz,  \textit{Detection of Gravitational Waves} in Proceedings of "Astrophysical Sources Of Gravitational Radiation" Editors:J. A. Marck and J. P. Lasota (Cambridge Univ. Press, 1996).
	
	\bibitem{Nowakowski} M. Nowakowski and I. Arraut, Acta Phys. Polon. B 41, 911 (2010).
\bibitem{Arraut} I. Arraut,  Mod. Phys. Lett. A 28, 1350019 (2013).	

\bibitem{Bicak} 	J. Bi\v{c}\'{a}k and J. Podolsk\'{y}, Phys. Rev. D 52, 887 (1995).
\bibitem{Tahamtan} T. Tahamtan and O Svitek, Phys. Rev. D 91, 104032 (2015).
\bibitem{Dereli} M. Adak, T. Dereli and Y. Senikoglu, Int. J. Mod. Phys. D. 28, 1950070 (2019). 

\bibitem{Bicak2} 	J. Bi\v{c}\'{a}k and J. Podolsk\'{y}, Phys. Rev. D 55, 1985 (1997).
\bibitem{Hatice} H. Ozer, PhD thesis, Istanbul University Physics Department, (2018).
\end{thebibliography}
\end{document}